\documentclass{aa}
\usepackage{graphicx}

\begin{document}

\title{On the atmospheric fragmentation of small asteroids}

\author{Luigi Foschini}

\offprints{L. Foschini, email: foschini@tesre.bo.cnr.it}

\institute{Institute TeSRE -- CNR, Via Gobetti 101, I-40129 Bologna,
Italy}

\date{Received 22 August 2000 / Accepted 16 October 2000}

\abstract{It is known, from observational data recorded from airbursts, that
small asteroids breakup at dynamical pressures lower than their
mechanical strength.  This means that actual theoretical models are
inconsistent with observations.  In this paper, we present a detailed
discussion about data recorded from airbursts and about several
theoretical models. We extend and improve a theory previously
outlined for the fragmentation of small asteroids in the Earth
atmosphere. The new condition for fragmentation is given by the
shock wave--turbulence interaction, which results in sudden outburst
of the dynamical pressure.  
}

\maketitle

\section{Introduction}
Collisions between cosmic bodies are divided into two regimes: 
gravity-- and strength--dominated regime.  But the transition from one 
regime to the other is not well known (see, for example, Fig.~5 in 
Durda et al.~1998).  In the case of collision with Earth, the 
presence of the atmosphere makes things more difficult.  Observations 
of airburst of small asteroids (up to tens of metres) show that the 
fragmentation occurs when the dynamical pressure is lower than the 
mechanical strength and this conundrum has not a satisfactory 
explanation yet (see Ceplecha~\cite{ZDENEK4}, Foschini~\cite{ME5}).  
It is worth noting that also the fracture itself has 
still many unknown features (for a review, see Fineberg \& Marder 
1999).

Studies on the fragmentation of small asteroids have also a great 
importance in impact hazard.  Although the damage caused by 
Tunguska--like objects can be defined as ``local'', it is not 
negligible.  The Tunguska event of 30 June 1908 resulted in the 
devastation of an area of $2150\pm 25$~km$^2$ and the destruction of 
more than 80 million trees (for a review, see Trayner~1997, Vasilyev~1998).  
Still today there is a wide debate all over the world about 
the nature of the cosmic body which caused that disaster.  Just on 
July 1999 an Italian scientific expedition, \emph{Tunguska99}, went to 
Siberia to collect data and samples (Longo et al.~\cite{LONGO}).

In this article, we discuss firstly the current models on atmospheric 
fragmentation of small asteroids and inconsistencies with observations 
(Sect.~2).  In Sect.~3, we deal with the question of strength and 
aerodynamic load.  In Sect.~4, we extend the approach previously 
outlined in Foschini~(\cite{ME3}), thereafter called Paper I. We study 
the hypersonic flow around a small asteroid, with particular reference 
to the definition of the pressure crushing on a cosmic body (Sect.~5).  
Therefore (Sect.~6), it is possible to obtain the condition for 
fragmentation under steady state conditions, that was already outlined 
in the Paper I. The Sect.~7 deals with the turbulence and massive 
ablation, putting the basis for the condition of fragmentation under 
unsteady conditions (Sect.~8).  Some numerical examples conclude the 
paper (Sect.~9).

\section{Problems with current models}
Present models consider that the fragmentation begins when the condition:

\begin{equation}
	\rho_{\infty}V^{2}=S
	\label{e:condfr}
\end{equation}

\noindent is satisfied. In Eq.~(\ref{e:condfr}), $\rho_{\infty}$ is the density 
of undisturbed stream, $V$ is the speed of the body, $S$ is the material 
mechanical strength. The term $\rho_{\infty}V^{2}$ is referred as the dynamical 
pressure in the front of the cosmic body.

However, observations of very bright bolides prove that large 
meteoroids or small asteroids breakup at dynamical pressures lower 
than their mechanical strength.  Today there is still no explanation 
for this conundrum.  This is a scientific problem of great interest, 
but it is also of paramount importance, because it allows us to know 
whether or not an asteroid might reach the Earth surface.  In addition 
to this, the atmospheric breakup has also effect on the crater field 
formation (Passey \& Melosh 1980) or on the area devastated by the 
airblast. 

The interaction of a cosmic body in the Earth atmosphere in the 
strength--dominated regime can be divided into two parts, according to 
the body dimensions.  For millimetre-- to metre--sized bodies 
(meteoroids), the most useful theoretical model is the 
gross--fragmentation model developed by Ceplecha et al.~(1993) and 
Ceplecha~(1999).  In this model, there are two basic fragmentation 
phenomena: \emph{continuous fragmentation}, which is the main process 
of the meteoroid ablation, and \emph{sudden fragmentation} or the 
discrete fragmentation at a certain point.

For small asteroids another model is used, where the ablation is 
contained in the form of explosive fragmentation, while at high 
atmospheric heights it is considered negligible.  Several models have 
been developed: Baldwin \& Shaeffer (1971), Grigoryan (1979), Chyba 
et al.  (1993), Hills \& Goda (1993), Lyne et al.  (1996).  A 
comparative study on models by Grigoryan, Hills \& Goda, and 
Chyba--Thomas-Zahnle was carried out by Bronshten (1995).  He notes 
that the model proposed by Chyba et al.  does not take into account 
fragmentation: therefore, the destruction heights are overestimated 
(about 10--12~km).  Bronshten also concludes that Grigoryan and 
Hills--Goda's models are equivalent.

Despite the particular features of each model, fragmentation is always 
considered to start when Eq.~(\ref{e:condfr}) is satisfied.

Although direct observations of asteroid impact are not available, it 
is possible to compare these models with observations of bodies with 
dimensions of several metres or tens of metres.  Indeed, in this 
range, the gross--fragmentation model overlaps the explosive 
fragmentation models.  As underlined several times by Ceplecha (1994, 
1995, 1996b), observations clearly show that meteoroids breakup at 
dynamical pressures lower (10 times and more) than their mechanical 
strength.  These data are obtained from photographic observation of 
bright bolides and the application of the gross--fragmentation model, 
that can be very precise.  

\begin{table}[t]
	\centering
	\caption{Meteoroid strength category.  After Ceplecha et al.  
	(1993)}
	\begin{tabular}{lrr}
	\hline
	Category & Range of $\rho_{\infty}V^{2}$ [MPa] & Mean
	$\rho_{\infty}V^{2}$ [MPa]\\
	\hline
	a & $\rho_{\infty}V^{2}<0.14$ & $0.08$\\
	b & $0.14 \leq \rho_{\infty}V^{2} < 0.39$ & $0.25$\\
	c & $0.39 \leq \rho_{\infty}V^{2} < 0.67$ & $0.53$\\
	d & $0.67 \leq \rho_{\infty}V^{2} < 0.97$ & $0.80$\\
	e & $0.97 \leq \rho_{\infty}V^{2} < 1.20$ & $1.10$\\
	\hline
	\end{tabular}
	\label{category}
\end{table}

According to Ceplecha et al.~(1993) it is 
possible to distinguish five strength categories with an average 
dynamical pressure of fragmentation (Table ~\ref{category}).
For continuous fragmentation the results obtained also indicate that 
the maximum dynamical pressure is below 1.2~MPa. Five exceptions 
were found: 4 bolides reached 1.5~MPa and one survived up to 5~MPa 
(Ceplecha et al. 1993).

It would be also very important to relate the ablation coefficient 
with the dynamical pressure $\rho_{\infty}V^{2}$ at the fragmentation 
point, in order to find a relationship between the meteoroid 
composition and its resistance to the air flow.  To our knowledge, a 
detailed statistical analysis on this subject does not exist, but in 
the paper by Ceplecha et al.  (1993) we can find a plot made by 
considering data on 30 bolides (we refer to Fig.~12 in that paper).  
We note that stony bodies (type I) have a wide range of 
$\rho_{\infty}V^{2}$ values at the fragmentation point.  In the case of 
weak bodies, we can see that there is only one cometary bolide (type 
IIIA), but this is due to two factors: firstly, cometary bodies 
undergo continuous fragmentation, rather than a discrete breakup at 
certain points.  Therefore, it is incorrect to speak about 
fragmentation pressure; we should use the maximum tolerable pressure.  
The second reason is that there is a selection effect.  Indeed, from 
statistical studies, Ceplecha et al.  (1997) found that a large part 
of bodies in the size range from 2 to 15~m are weak cometary bodies.  
However, a recent paper has shown that statistics from physical 
properties can lead to different results when compared with statistics 
from orbital evolution (Foschini et al.~2000).  To be more precise, 
physical parameters prove that, as indicated above, a large part of 
small near Earth objects are weak cometary bodies, whilst the analysis 
of orbital evolution proves a strong asteroidal component.

In addition to data published in the paper by Ceplecha et al.~(1993) 
and Ceplecha (1994) we can see in Table ~\ref{special} some specific cases
of bright bolides; for details, see the papers quoted or to 
Foschini~(\cite{ME5}).

\begin{table*}[ht]
\centering
\caption{Some special episodes of superbolides. Pressures are expressed in MPa.}
\begin{tabular}{lrrrr}
\hline
Name & Date & $\rho_{\infty}V^{2}$ & $S$ & Source\\
\hline
P\v{r}\`{\i}bram & Apr 7, 1959 & 9.2 & 50 & ReVelle~(1979)\\
Lost City & Jan 3, 1970 & 1.5 & 50 & Ceplecha~(1996a)\\
\v{S}umava & Dec 4, 1974 & 0.14 & 1 & Borovi\v{c}ka \& Spurn\'y~(1996)\\
Innisfree & Feb 6, 1977 & 1.8 & 10 & ReVelle~(1979)\\
Space based obs.  & Apr 15, 1988 & 2.0 & 50 & Nemtchinov et al.~(1997)\\
Space based obs.  & Oct 1, 1990 & 1.5 & 50 & Nemtchinov et al.~(1997)\\
Bene\v{s}ov & May 7, 1991 & 0.5 & 10 & Borovi\v{c}ka et al.~(1998a, b)\\
Peekskill & Oct 9, 1992 & 1.0 & 30 & Ceplecha et al.~(1996)\\
Marshall Isl.  & Feb 1, 1994 & 15 & 200 & Nemtchinov et al.~(1997)\\
\hline
\end{tabular}
\label{special}
\end{table*}

\section{Stresses and strengths}
For the sake of the simplicity, in the above section, we adopted
commonly used values for mechanical strength, i.e.  1~MPa for cometary 
bodies, 10~MPa for carbonaceous chondrites, 50~MPa for stony bodies, 
and 200~MPa for iron bodies (see, for example, Hills \& Goda 
\cite{HILLS}).  Only Bronshten (\cite{BRON2}) rejected the values for 
cometary bodies, proposing the range $0.02-0.4$~MPa. He used values 
calculated by \"Opik (\cite{OPIK}) after observation of tidal 
disruption of Sun--grazing comets under the gravity of the Sun. 
However, it is worth noting that the strength obeys to scaling laws: 
therefore the larger is the body, the smaller is the strength. 
\"Opik's calculations refer to the comet Ikeya--Seki (1965f), which 
has an estimated nucleus, before the breakup, of about 8.3~km. Small 
asteroids are in the range of several tens of metres, up to some 
hundreds of metres. The Tunguska Cosmic Body (TCB), to which Bronsthen 
apply his calculations, is in the range $50-100$~m. We can calculate 
how the scaling law changes the value of the strength. The formula is 
shown in the paper by Tsvetkov \& Skripnik (\cite{TSVETKOV}) and we 
recall it for simplicity:

\begin{equation}
	S=S^{\prime}\left(\frac{m^{\prime}}{m}\right)^\alpha
	\label{e:scaling}
\end{equation}

\noindent where $m$ and $m^{\prime}$ are the masses of the bodies and 
$\alpha$ is the scale factor: the more inhomogeneous is the material, 
the higher is $\alpha$. Turcotte (\cite{TURCOTTE}) adopted a value 
of $\alpha=0.12$ for glacial tills and we use this value. The comet 
mass can be calculated as $3\cdot10^{14}$~kg, by using a density of 
$1000$~kg/m$^{3}$, while the Tunguska Cosmic Body has an estimated 
value around $10^{8}$~kg (Vasilyev \cite{VASI}). With these values in 
Eq.~(\ref{e:scaling}) we obtain a value of the mechanical strength for 
a Tunguska--sized body, in the range $0.12-2.4$~MPa, that is compatible 
with commonly used value of 1~MPa. 

The scaling law for mechanical strength derives from the assumption 
that the fragmentation was a process of consecutive elimination of 
defects under increasing load. Baldwin \& Sheaffer (\cite{BALDWIN}) 
consider that the reason for the presence of cosmic bodies with very low 
fragmentation pressure can be explained by the assumption that 
additional flaws and cracks may be created by collisions in space, 
even though they do not completely destroy the cosmic body. 

Tsvetkov \& Skripnik (\cite{TSVETKOV}) made an interesting study on the
fragmentation according to strength theory and scaling laws, but this 
kind of study is useful only for meteorites and cannot explain 
airbursts. They also show that the condition of fragmentation 
$S=\rho_{\infty}V^{2}$ is not valid even under the assumptions of scaling 
laws. Indeed, they showed that the aerodynamic loading never reach the 
ultimate strength. For this reason, they searched the cause of 
fragmentation in the particular structure and extreme inhomogeneity of 
cosmic bodies. With an appropriate selection of the scale factor, Tsvtkov 
\& Skripnik obtained a good fit. 

However, the shock compression and heating during the atmospheric 
entry will result in an elimination of internal cracks, making the 
body more compact (see, for example, Davison \& Graham 1979 or 
Zel'dovich \& Raizer 1966).  If internal voids are so large to
survive to shock compression, they could give an explanation for some 
episodes, but not a general theory (Foschini 1998).

\section{The hypersonic flow approach}
Almost all models described in Sect.~2 deal with the 
motion of a cosmic body in the Earth atmosphere.  However, it is 
worth noting that we cannot observe directly the cosmic body: we can 
only see the light emitted during its atmospheric entry.  Therefore, 
we have to introduce in the equations several coefficients, that cannot be 
derived directly from observations.

If we turn our attention to the hypersonic flow around the body, we 
could have data from direct observations.  Among models discussed 
above, only Nemtchinov et al.  (1997) tried to investigate the 
hypersonic flow around the asteroid with a numerical model.  Foschini 
(1999b) investigated the analytic approach: indeed, although the 
details of hypersonic flow are very difficult to calculate and 
there is need of numerical models, the pressure distribution can be 
evaluated with reasonable precision by means of approximate methods.  
In the limit of a strong shock ($M>>1$) several equations tend to 
asymptotic values and calculations become easier.  For example, the 
ratio of densities across the shock is:

\begin{equation}
	\frac{\rho_{0}}{\rho_{\infty}}\rightarrow \frac{\gamma+1}{\gamma-1}
	\label{e:density}
\end{equation}

\noindent where $\rho_{0}$ is the density in the stagnation point and 
$\gamma$ is the specific heats ratio.

In the Paper I, it was showed the crucial role of the temperature, 
instead of the pressure, in the stagnation point.  The gas is in local 
thermodynamic equilibrium (LTE), that is, matter is in equilibrium 
with itself, but not with radiation, which can escape.  Particle 
densities depend on the temperature only and it is possible to use 
Boltzmann or Saha equations.  The temperature is calculated directly 
from observations of spectra (e.g.  Borovi\v{c}ka \& Spurn\'y 1996, 
Borovi\v{c}ka et al.  1998b) and, thus, it is a reliable starting point.  

Particularly, as stated in the Paper I, the temperature at the stagnation 
point is very important.  Changes in the stream properties are mainly due 
to changes in the stagnation temperature, which is a direct measure of 
the amount of the heat transfer.  The enthalpy change $\Delta h$ is:

\begin{equation}
	\Delta h = c_{\mathrm{P}}\Delta T
	\label{e:entalpia}
\end{equation}

\noindent where $c_{\mathrm{P}}$ is the specific heat with constant pressure.  
From Eq.~(\ref{e:entalpia}), it is possible to relate the maximum 
speed of the stream (which is close to the cosmic body speed) to the 
stagnation temperature (see Paper I):

\begin{equation}
	V_{\mathrm{max}}=\sqrt{\frac{2\gamma}{\gamma-1}RT_{0}}
	\label{e:speed}
\end{equation}

\noindent where $R$ is the constant of the specific gas.  The 
Eq.~(\ref{e:speed}) is valid for an adiabatic flow whether reversible 
or not.

Under certain approximations, the flow can be considered adiabatic: 
heat transfer in hypersonic motion is strongly reduced.  The boundary 
layer, where convective heat transfer takes place, is very thin and 
can be considered negligible.

On the other hand, if we consider the Eq.~(\ref{e:speed}), we see that 
it is satisfied for unrealistic high temperatures.  For example, for a 
speed of $V=12$~km/s in air we obtain $T_{0}=72000$~K. The 
introduction of ablated matter make $\gamma$ higher and $R$ lower, 
resulting in a further increase of $T_{0}$.

However, at high temperatures some effects of the gas become 
important: the main effects are the dissociation and ionization of air 
molecules.  For $2000<T<4000$~K the oxygen molecules break down to 
single atoms; for $4000<T<9000$~K there is the dissociation of 
$N_{2}$; for $T>9000$~K ionization occurs (see, for example, 
Oosthuizen \& Carscallen 1997).  In the $\Delta h$ term of the 
Eq.~(\ref{e:entalpia}), we should take into account also the heat 
absorbed by these processes.  As a result, $T_{0}$ drops to 
more reasonable $10000-15000$~K or so.

The treatment of real gas properties is probably one of the most 
difficult problem in aerothermodynamics and involves complex numerical 
modelling (for a review, see Tirsky 1993, Gnoffo 1999).  

\section{The equation of state}
Before going on, it is necessary to evaluate the state of the shocked 
gas, in order to select the more appropriate equation of state.

From the point of view of pressure, we can see that the gas around an 
asteroid during the atmospheric entry, is still an ideal gas.  Indeed, the 
limit of pressure to become a degenerate gas is around $10^{14}$~Pa 
(Eliezer 1991).  A simple calculation shows that the asteroid reaches 
the maximum dynamical pressure at the sea level 
($\rho_{\mathrm{sl}}=1.293$~kg/m$^{3}$) and if it has the maximum geocentric 
speed ($V=72$~km/s, although it is quite impossible to have an 
asteroid with such a speed).  Therefore, the maximum 
$\rho_{\infty}V^{2}\approx 7\cdot 10^{9}$~Pa, about four orders of 
magnitude below the limit of degeneracy. 

From the point of view of the temperature, the question is a bit 
difficult to handle.  From spectroscopic observations of bright 
bolides (Borovi\v{c}ka \& Spurn\'y 1996, Borovi\v{c}ka et al.  
1998b), we can see that the spectrum is composed by two temperatures 
emission: the first value lies in the range between $4000-6000$~K and 
the second value is about $10000$~K. The temperature of $10000$~K is 
about the limit between normal and partially ionized gas. Therefore, we shall
adopt a more general point of view, based on microscopic physics.

From the kinetic theory, we know that the pressure is additive, therefore
we can separate it in contributions from different species composing the gas:
electrons, photons, ions, and atoms. We can write:

\begin{equation}
P=P_{\mathrm{i}}+P_{\mathrm{e}}+P_{\mathrm{r}}=
\sum_{\mathrm{j}}N_{\mathrm{j}}kT_{\mathrm{j}}+N_{\mathrm{e}}kT_{\mathrm{e}}+\frac{aT^{4}}{3}
\label{e:pressione4}
\end{equation}

\noindent where $k$ is the Boltzmann constant and $a$ is the radiation
constant, related to the Stefan--Boltzmann constant
($a=4\sigma/c=7.56591\cdot 10^{-16} [\mathrm{Pa\cdot K^{-4}}]$);
$N$ is the volume density of species (the index $e$ is for electrons);
the sum over the index $j$ concerns all heavy particles (ions and neutral
atoms).

Eq.~(\ref{e:pressione4}) has been obtained by means of quite general
assumptions, without any particular restrictions (more details can be found in
many places; for example, in Lang 1999).  We can apply
Eq.~(\ref{e:pressione4}) to hypersonic flow around an asteroid in the Earth
atmosphere.  The only restrictions are determined by the momentum
distributions of particles: ions, atoms, and electrons obey generally to a
Maxwellian distribution, photons obey to the Bose--Einstein statistics. These
conditions are satisfied in processes occurring during the atmospheric entry
of a cosmic body.

\section{The pressure of fragmentation}
We can note that, for typical temperatures in airbursts, the radiation 
pressure is negligible.  For $T=10000$~K, we have that $P_{\mathrm{r}}\approx 
2.5$~Pa, which is negligible when compared to $\rho_{\infty}V^{2}$ at the 
fragmentation point (see Sect.~2).  Therefore, for airbursts and 
meteors, the Eq.~(\ref{e:pressione4}) becomes:

\begin{equation}
P\approx P_{\mathrm{i}}+P_{\mathrm{e}}=
\sum_{\mathrm{j}}N_{\mathrm{j}}kT_{\mathrm{j}}+N_{\mathrm{e}}kT_{\mathrm{e}}
\label{e:pressione5}
\end{equation}

In order to calculate the densities of species, we have to take into 
account that the fluid around the asteroid is in local thermodynamic 
equilibrium (LTE) and we can use the Saha relation:

\begin{equation}
	\frac{N_{\mathrm{e}}N_{\mathrm{i}}}{N_{\mathrm{a}}}=\frac{(2\pi 
	mkT)^{3/2}}{h^{3}}\frac{2g_{\mathrm{i}}}{g_{\mathrm{a}}}\exp(-\frac{eE_{\mathrm{i}}}{kT})
	\label{e:saha}
\end{equation}

\noindent where $N_{\mathrm{a}}$ is the density of neutral atoms, $E_{\mathrm{i}}$ is 
the ionization potential, $g_{\mathrm{a}}$ and $g_{\mathrm{i}}$ are respectively the 
statistical weight of the ground state of the neutral atom and of the 
ion.  Some of these values are listed in Table ~\ref{species}.

\begin{table}[ht]
\centering
\caption{Values of $E_{\mathrm{i}}$, $g_{\mathrm{a}}$, and $g_{\mathrm{i}}$ for some species in 
the flow}
\begin{tabular}{lrrr}
\hline
Species & $E_{\mathrm{i}}$ [eV] & $g_{\mathrm{a}}$ & $g_{\mathrm{i}}$\\
\hline
Na & 5.14 & 2 & 1\\
K & 4.34 & 2 & 1\\
O$_{2}$ & 12.05 & 3 & 4\\
N$_{2}$ & 15.6 & 1 & 2\\
\hline
\end{tabular}
\label{species}
\end{table}

For each airburst, with a given composition, we can calculate 
densities of all species and obtain the effective pressure.  But this 
is not the purpose of this work.  We will introduce some hypoteses for 
the sake of the simplicity, in order to do some calculations of order 
of magnitude.

\begin{figure*}[!ht]
\centering
\includegraphics[angle=270,scale=0.6]{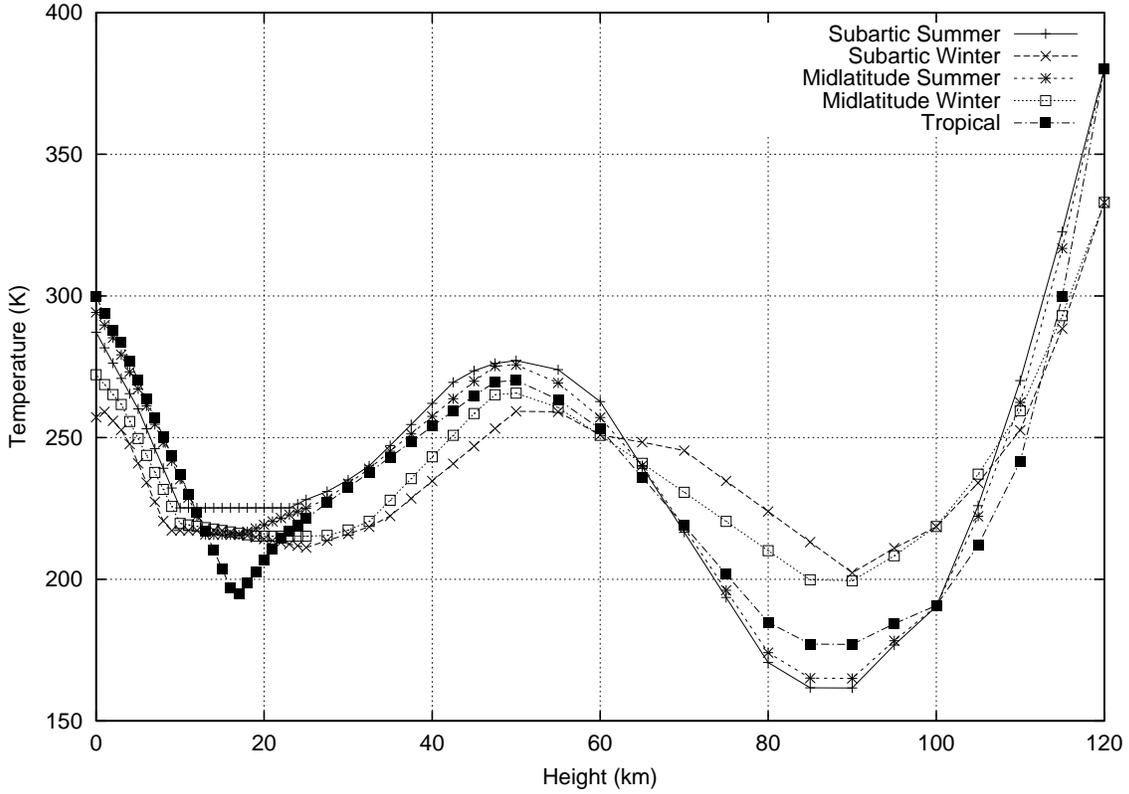}
\caption{Temperature as a function of the height, for different latitudes and 
seasons. Data from US Standard Atmosphere 1976.}
\label{FIG1}
\end{figure*}

The Saha equation (\ref{e:saha}) can be rewritten in terms of degree 
of ionization $\alpha$:

\begin{equation}
	\frac{N_{\mathrm{tot}}\alpha^{2}}{(1-\alpha)}=\frac{(2\pi 
	mkT)^{3/2}}{h^{3}}\frac{2g_{\mathrm{i}}}{g_{\mathrm{a}}}\exp(-\frac{eE_{\mathrm{i}}}{kT})
	\label{e:saha2}
\end{equation}

\noindent where $N_{\mathrm{tot}}$ is the total number density of particles; 
therefore, $N_{\mathrm{e}}=\alpha N_{\mathrm{tot}}$ and so on. The 
Eq.~(\ref{e:pressione5}) can be rewritten:

\begin{equation}
	P=(1+\alpha)N_{\mathrm{tot}}kT=(1+\alpha)\rho RT
	\label{e:pressione6}
\end{equation}

\noindent that is the well known equation of state for a gas with a 
degree of ionization $\alpha$. 

Foschini (1999a) underlined the importance of alkaline metals in the 
production of electrons in meteors.  In addition, von Zahn et al.  
(\cite{ED}) recently showed the particular role of the potassium in 
Leonid meteors.  Indeed, alkaline metals, owing to their low 
ionization potential easily ionize even at low temperatures.  For 
example, we assume that $N_{\mathrm{tot}}=6.022\cdot 10^{23}$~m$^{-3}$ (it is 
the Avogadro number) and we calculate the degree of ionization of a 
gas composed by single species listed in Table ~\ref{species}.  For a
temperature $T=10000$~K, we have: for sodium, $\alpha=0.92$; for 
potassium, $\alpha=0.96$; for oxygen, $\alpha=0.1$ and for nitrogen, 
$\alpha=0.01$.

It is known that, in meteors, the impact ionization from impinging 
molecules is very important.  A nitrogen molecule with a speed of 
18~km/s has an energy of about 45~eV. However, as the asteroid 
penetrates deeper in the atmosphere, the flow generates a shield that 
protects the asteroid from direct impacts (for a detailed description 
see Ceplecha et al.~1998).  Therefore, the thermal ionization become 
important particularly in small asteroids/comets, when impact 
ionization is strongly reduced.  So, alkaline metals play an important 
role, because they provide a huge amount of electrons.

We have also to take into account that we are dealing with particular 
condition in the stagnation point, where there is the maximum 
thermo--mechanical stress. In this point, we can consider the fluid 
as highly ionized: the ablation of alkaline metals provide a source 
of ions for the flow. In the stagnation point -- not in the whole 
flow -- we can consider the gas with a degree of ionization $\alpha \rightarrow 
1$. Therefore, by substituting Eq.~(\ref{e:pressione6}) in 
Eq.~(\ref{e:speed}), we obtain the condition for fragmentation:

\begin{equation}
	V_{\mathrm{max}}=\sqrt{\frac{2\gamma}{\gamma-1}\frac{P}{(1+\alpha)\rho}}
	\label{e:speed2}
\end{equation}

Eq.~(\ref{e:speed2}) was already outlined in the Paper I, 
where it was considered $\alpha=1$. 

It is worth noting that Eq.~(\ref{e:speed2}) is the condition of 
fragmentation, not the condition for the airburst, as deduced by 
Bronshten (\cite{BRON2}). The airburst generally occurs after a scale 
height, as shown by several studies on superbolides. In the Paper I, there 
was an unfortunate error in considering the explosion height, instead 
of the fragmentation height (although it has negligible effect, as we 
shall see in the Sect.~9). 
Therefore, Eq.~(\ref{e:speed2}) must be revised as follows:

\begin{equation}
	V_{\mathrm{max}}=\sqrt{\frac{2\gamma}{\gamma-1}\frac{P}{(1+\alpha)\rho_{\mathrm{sl}}}
	\exp(\frac{h+H}{H})}
	\label{e:speed3}
\end{equation}

\noindent where $\rho_{\mathrm{sl}}$ is the air density at sea level, $h$ is 
the airburst height, and $H$ is the scale height. 

Before to analyse the Eq.~(\ref{e:speed3}), it is necessary to look at 
the value for $\gamma$. In the Paper I, $\gamma=1.7$ was used and it was 
taken from experimental measurement in hypervelocity impacts. 
Bronshten (\cite{BRON2}) does not agree with 
this value and invokes argument to support a value of $\gamma=1.15$, 
that is compatible with air at high temperatures and under shock 
loading. Bronshten's assumption are reasonable when we have a diatomic 
gas at high temperature: the rotation and vibration of air molecules 
change the values for $\gamma$. But, if we have a monatomic gas or 
metal vapors, we have to consider a $\gamma=5/3$, unless the temperature 
is so high to change appreciably the electron energy 
(see Zel'dovich \& Raizier \cite{ZELDO}); 
however, this is not our case. This is confirmed also by meteor spectra: 
they show that the surrounding gas is monatomic, although molecules bands are 
sometime recorded (Ceplecha et al. \cite{ZDENEK9}).

It is worth noting that the question about the value of $\gamma$ is not
simple. Perhaps, it will be solved when experimental data will be available.

\begin{figure*}[!ht]
\centering
\includegraphics[angle=270,scale=0.6]{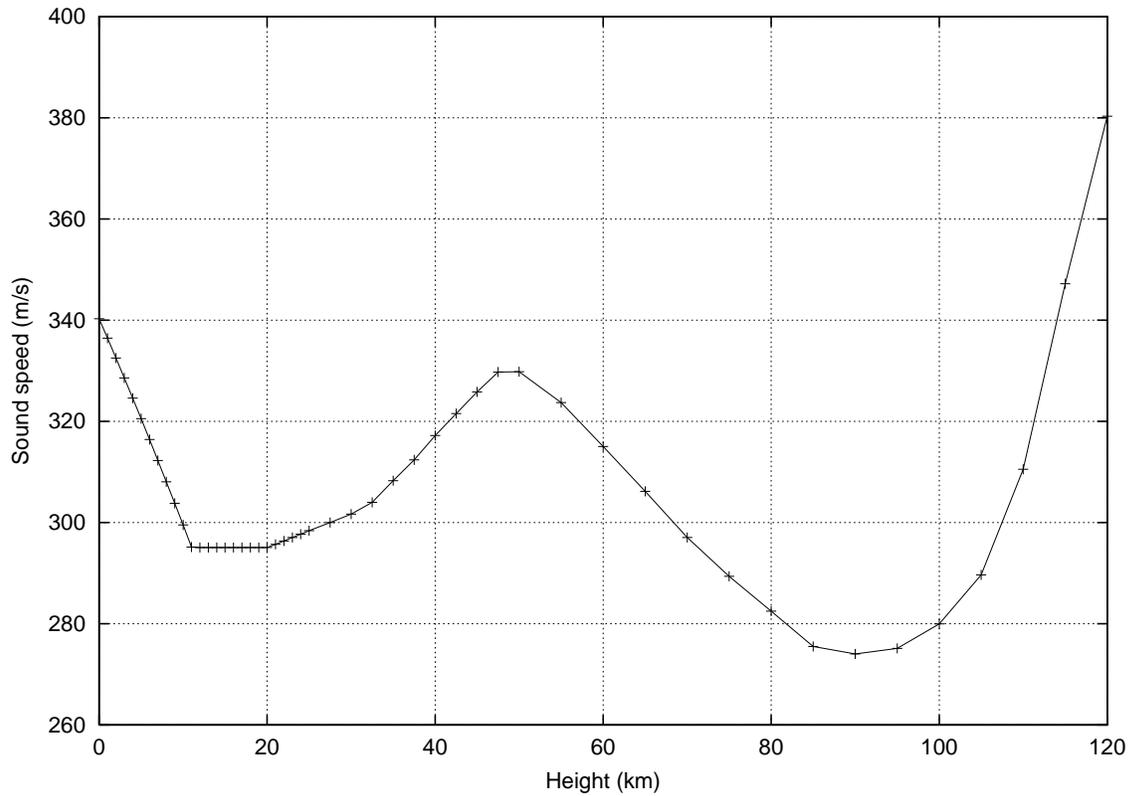}
\caption{Local sound speed from the reference temperature of the 
US Standard Atmosphere 1976.}
\label{FIG2}
\end{figure*}

\section{Turbulence and massive ablation}
Eq.~(\ref{e:speed3}) shows that it is necessary to have higher speeds 
in order to reach the value of the mechanical strength. It is worth 
noting that Eq.~(\ref{e:speed3}) is valid under steady state condition. 
In this case, we can neglect the contribution of turbulence, because 
of compressibility effects at high Mach number (see Andreopulos et 
al.~\cite{TURBO1}).

The influence of the turbulence in the large meteoroid entry was first 
analyzed by ReVelle~(\cite{REV}).  In his approach based on the motion 
of a single body in the atmosphere, ReVelle studied how the convective 
heat transfer depends on turbulence.  He found that the turbulent 
convective layer is negligible, except for speed lower than about 
$20$~km/s.

However, the presence of massive ablation changes the flow.  
Gupta~(\cite{GUPTA}) noted that there are few experimental data 
available and, most important, they are obtained with freestream Mach 
number in the range 3--7 and with negligible ablation.  There are less 
data about experiments with moderate ablation, but with lower Mach 
number (up to 2.6).  Starting from these data, models predict that 
convective heating is negligible with almost any turbulence model.  
This was the situation in 1983, but more recent reviews 
(Gnoffo~\cite{GNOFFO}) show that there are still several things 
unknown.  The only news is the entry of the Galileo probe in the 
Jovian atmosphere, which gave us useful information about hypersonic 
motion with massive ablation.  The probe entered in the atmosphere at 
a relative speed of $47.5$~km/s and experienced an ablation rate of 
$7.4$~km/s, with a total mass loss of about $79$~kg 
(Gnoffo~\cite{GNOFFO}).

The probe did not suffer any fragmentation, despite of massive 
ablation, although it is still not clear the role of all processes 
during the entry.  The layer between the shock and the probe is 
expected to be turbulent over almost the entire length of the Galileo 
probe, owing to the massive ablation and the large Reynolds number 
(Gupta~\cite{GUPTA}).  The interaction of shock waves with turbulence 
leads to amplification of speed fluctuations and changes in length 
scales.  It would be better to say that this occur when the flux is 
\emph{unsteady} and the shock wave is subject to strong distortions.  
At high Mach number, but steady state motion, the effect of 
compressibility takes place and there is no amplification; we can 
adopt Eq.~(\ref{e:speed3}).

But the atmospheric motion is generally unsteady.  Fig.~\ref{FIG1} shows the air 
temperature from tables of the US Standard Atmosphere 1976.  The air 
temperature is instrumental to derive the local sound speed 
($a=\sqrt{\gamma RT}$), as a function of the height, for different 
latitude and seasons.  As the temperature changes, the sound speed 
changes.  The US Standard Atmosphere gives also a reference atmosphere 
-- and therefore a reference temperature -- from which we can 
calculate a reference sound speed (Fig.~\ref{FIG2}).

\begin{figure*}[!ht]
\centering
\includegraphics[angle=270,scale=0.6]{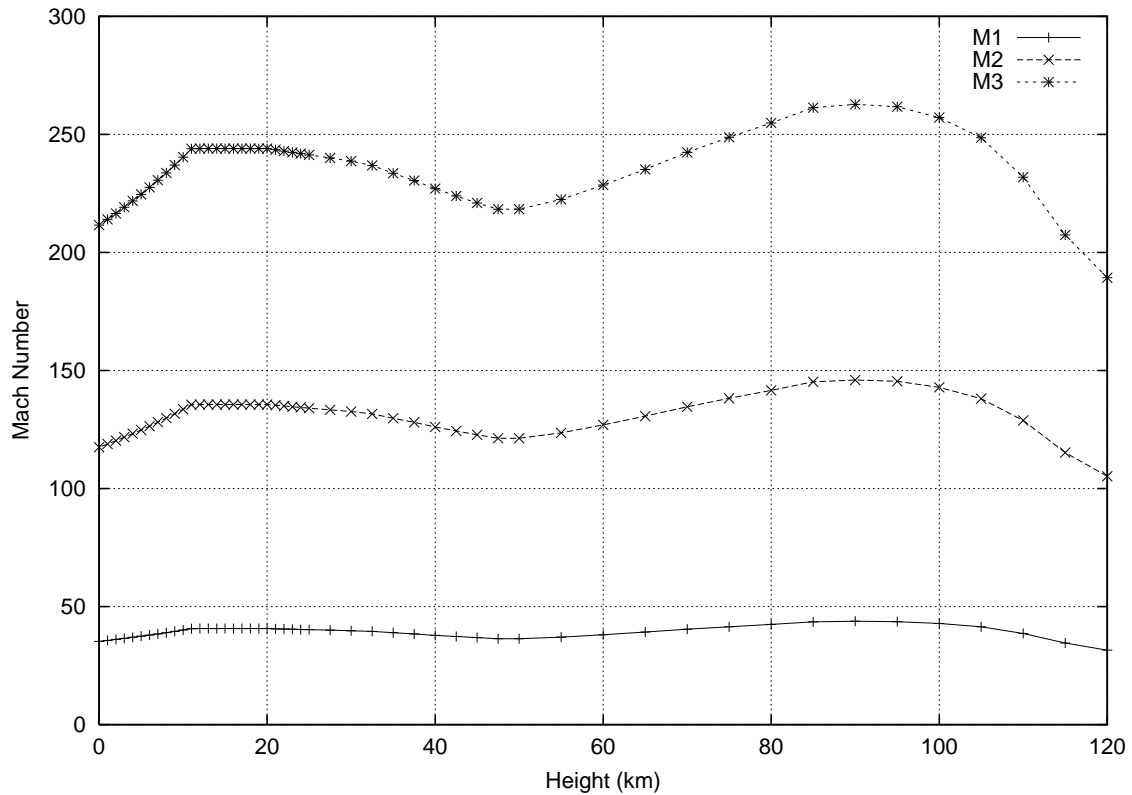}
\caption{Mach number as a function of the height for different asteroid speed. 
M1=12~km/s; M2=40~km/s; M3=72~km/s.}
\label{FIG3}
\end{figure*}

It is known that, for a small asteroid, the deceleration due to 
ablation is negligible; therefore we can consider that the body speed 
does not change until the fragmentation begins.  For a given asteroid, 
the Mach number depends only on the local sound speed.  In 
Fig.~\ref{FIG3} we can see the Mach number for different values of the 
speed of the cosmic body.  It results that the Mach number change substantially 
for higher asteroid speed, while for lower speed, changes of M become 
smooth.  We note also strong variations when the body crosses the 
mesopause, the stratopause, and the tropopause.

Ceplecha (\cite{ZDENEK1}) gave the mean values of end heights for 
large meteoroids (up to about 7~m): he found 32~km for type I; 43~km 
for type II; 57~km for type IIIA; and 69~km for type IIIB. These are 
end heights and if we consider that the fragmentation begins about a 
scale height above, we infer that the mean fragmentation height is: 
39~km for type I; 50~km for type II; 64~km for type IIIA; and 77~km 
for type IIIB. That is, we can divide these bodies into two 
categories: a first category, made with type II and type IIIB, which 
contains bodies that breakup during the crossing of the stratopause 
and the mesopause, respectively.  The second category, made with type 
I and type IIIA, which contains bodies that breakup with a certain 
delay after the crossing of atmospheric pauses.  The delay can be 
explained with the different ability to ablate, and therefore to 
produce the turbulent boundary layer.  Indeed, for bodies that breakup 
at the mesosphere (cometary bodies), the type IIIB has an average 
ablation coefficient higher than the type IIIA. For asteroidal bodies, 
that breakup at the stratosphere, the type II has a higher ablation 
coefficient than the type I. We should also take into account that the 
transition from laminar to turbulent flow depends on the body 
dimensions and speed.  Large bodies at high speed reach the transition 
at higher heights (ReVelle~\cite{REV}).

It is worth noting that there are also other factors that can change 
the fragmentation height, such as the rotation (Adolfsson \& 
Gustafson~\cite{ADOLF}, Ceplecha~\cite{ZDENEK3}), but they are 
independent from the body type and can explain only deviations from the 
average values.

\section{The condition for fragmentation}
To find a quantitative condition for fragmentation by taking into 
account the turbulence, at least for first order calculations, is very 
difficult.  Problems derive from the lack of experimental data about 
the shock wave--turbulence interaction at very high Mach number.  
Moreover, the few feasible experiments at moderate Mach number are 
strongly dependent on measurement systems and available data are 
insufficient to fully describe the mutual interaction of shock waves 
and turbulence.  Last, but not least, the turbulence itself is 
one of the oldest unsolved problems in the history of science.
In recent years, numerical models allowed detailed 
investigation, but the problem is complicated by the fact that, when 
dealing with turbulence, the averaging of governing equations 
introduces new unknowns. Therefore, the number of available equations 
is not sufficient and it is necessary to assume a closure condition. 
The lacking of experimental data make it hard to make hypoteses on the closure 
condition and therefore numerical models are often contradictory. For 
reviews of these problems see Andreopulos et al.~(\cite{TURBO1}), 
Lele~(\cite{LELE}), Adamson \& Messiter~(\cite{TURBO2}), and 
references therein. 

Despite of differences, it is clear that the shock wave--turbulence 
interaction produce an amplification of fluctuations. The 
amplification depends on the shock strength, the state of the 
turbulence, and its level of compressibility. The most important 
outcomes are the amplification of velocity fluctuations and changes in 
the length scales (Andreopulos et al.~\cite{TURBO1}). This leads to 
changes in the dynamical pressure in the front of the asteroid, but 
also changes in pressure along the flank of the cosmic body; these 
changes can be further amplified by local irregularity, such as small 
``hills''. 

However, for the sake of simplicity, we make the assumption that the 
zone of greatest stress is the stagnation point. Or, it would be 
better to say, that we will continue to hold this hypotesis. Being 
under unsteady conditions we cannot apply isentropic relations, i.e. 
Eq.~(\ref{e:speed3}). We start from the value of the pressure in the 
shock layer derived from the Rankine--Hugoniot relations in the 
hypersonic limit ($P\approx \rho_{\infty}V^{2}$). Being the gas cap 
around the asteroid a plasma, we have to consider the 
Eq.~(\ref{e:pressione6}), so we have:

\begin{equation}
	P\approx (1+\alpha)\rho_{\infty}V^{2}
	\label{e:fragm1}
\end{equation}

For the sake of the simplicity, we assume that turbulence does not 
affect density; so we can consider that shock wave--turbulence 
interaction does affect only velocity. The problem is to set up the 
value of this amplification: we then introduce an amplification 
factor $\kappa$ to evaluate:

\begin{equation}
	P\approx (1+\alpha)\rho_{\infty}\kappa V^{2}
	\label{e:fragm2}
\end{equation}

As written above, we have no experimental data and numerical models 
are often contradictory.  We can try to set up lower and upper limits.  
Rotman~(\cite{TURBO4}) reports amplification of the kinetic energy of 
about $2-2.15$.  Jacquin et al.~(\cite{TURBO3}) found that the 
amplification of the kinetic energy depends on the density ratio; the 
factor can be up to $12.7$ for diatomic gases.  For monatomic gases 
and plasmas the upper limit of the amplification is $6$. 

In conclusion, we can assume for kinetic energy $2 \leq\kappa\leq 6$ and that 
this amplification value is valid also for pressure. Therefore, under 
distortion of the shock wave, the turbulence can leads to amplification 
of dynamical pressure up to 12 times the nominal value for a neutral 
gas (we have taken into account also the multiplicative factor $1+\alpha$). 
Comparing with experimental data of the fragmentation of asteroids 
showed in Sect.~2, we can see that they are in better agreement (see 
Table ~\ref{fragnew}).

\begin{table}[ht]
\centering
\caption{Special episodes of superbolides: the new pressure of 
fragmentation calculated according to Eq.~(\ref{e:fragm2}). See Table ~2
for other details. Pressures are expressed in MPa.}
\begin{tabular}{lrrr}
\hline
Name & Min $P$ & Max $P$ & $S$\\
\hline
P\v{r}\`{\i}bram & 37 & 110 & 50 \\
Lost City & 6 & 18 & 50 \\
\v{S}umava & 0.6 & 1.7 & 1 \\
Innisfree & 7 & 22 & 10 \\
Space based obs.  & 8 & 24 & 50 \\
Space based obs.  & 6 & 18 & 50 \\
Bene\v{s}ov & 2 & 6 & 10 \\
Peekskill & 4 & 12 & 30 \\
Marshall Isl.  & 60 & 180 & 200 \\
\hline
\end{tabular}
\label{fragnew}
\end{table}

Therefore, the new condition for fragmentation under unsteady regime is:

\begin{equation}
	V=\sqrt{\frac{S}{\kappa (1+\alpha) \rho_{\mathrm{sl}}}\exp(\frac{h+H}{H})}
	\label{e:fragm3}
\end{equation}

In addition, we have to consider that the distortion of the shock wave leads to
the partial removal of the gas cap around the cosmic body, so that the ablation
increases strongly. This enhances impact ionization, so that 
$\alpha\rightarrow 1$ also during unsteady conditions, where thermal 
ionization is negligible. 

\section{Examples}
Let us to consider two episodes in order to show how this ``embryo'' of 
theory works. Firstly, we can consider the Lugo bolide of 19 January 
1993 (Cevolani et al.~\cite{CEVO}, Foschini~\cite{ME1}). It was a 
very bright bolide, which reached a peak magnitude of about $-23$ and 
released an estimated energy of about 14~kton, when exploded at about 
30~km over the city of Lugo, in northern Italy. In previous analyses, 
it was considered that the fragmentation occurred when the dynamical 
pressure reached the mechanical strength. 

If we now apply the condition given from Eq.~(\ref{e:fragm3}), taking 
into account $\alpha \rightarrow 1$, $\rho_{\mathrm{sl}}=1.293$~kg/m$^{3}$, 
and that the scale height $H=6.8$~km at 30~km height, we obtain the 
values in Table ~\ref{TAB4}.

\begin{table}[ht]
\centering
\caption{Values of the speed of the Lugo bolide at the moment of 
fragmentation [km/s].}
\begin{tabular}{lcc}
\hline
S [MPa] & $\kappa=2$ & $\kappa=6$\\
\hline
1 & 6.6 & 3.8\\
10 & \underline{20.8} & \underline{12.0}\\
50 & 46.5 & \underline{26.9}\\
200 & 93.1 & 53.7\\
\hline
\end{tabular}
\label{TAB4}
\end{table}

We have three reasonable solutions, which are underlined. If we 
consider that the final airburst occurred at about 30~km height, 
typical for type I bodies (see the discussion at the end of Sect.~7), 
we can consider 26.9~km/s as the most probable speed. This value is in 
agreement with first estimation from eyewitnesses (Cevolani et 
al.~\cite{CEVO}).   

For the Tunguska event, a more detailed analysis is in preparation in 
collaboration with the members of the Tunguska99 Scientific 
Expedition.  Here we want to underline only one thing: from 
Fig.~\ref{FIG1} we can see that, in subartic summer, the temperature 
does not change in the height interval crossing the troposphere.  
Therefore, the Mach number does not change and we can apply the 
Eq.~(\ref{e:speed3}), for steady state conditions, as shown in the Paper I.  
The error, noted by Bronshten~(\cite{BRON2}), 
introduced in considering the explosion height, instead of the 
fragmentation height is negligible: the new value is 16~km/s to be 
compared with the old one of 16.5~km/s.

\section{Conclusion}
In this paper, we have done a further step toward the construction of 
a theory for the fragmentation of a small asteroid during the 
atmospheric entry. In the Paper I, we showed a specific part of the 
theory applied to a particular episode, the Tunguska event of 30 June 1908. 

Here we showed more details, both explaining better some assumptions 
in the theory of steady state motion outlined in the Paper I, 
and extending the theory to unsteady motion.  We have taken into 
account the effect of turbulence and its interaction with shock waves.  
Some examples are discussed and we have found a reasonable agreement 
with available experimental data.

We have found two conditions for the fragmentation, according to steadiness
of the motion.

\begin{enumerate}

\item Steady state motion: in this case, the compressibility suppresses the 
turbulence and, therefore, we can use Eq.~(\ref{e:speed3}). 

\item Unsteady motion: we have a strong interaction between the shock wave and
the turbulence, which give rise to sudden pressure outburst. We have to use 
Eq.~(\ref{e:fragm3}). 

\end{enumerate}

On the other hand, it is necessary to remember that we are 
speculating, because of the scarce experimental data and the large 
uncertainties affecting records. Therefore, these researches 
must be taken \emph{cum grano salis}.

\begin{acknowledgements}
I wish to thank R.~Guzzi for providing the computer--ready files with
data of US Standard Atmosphere 1976, and G.~Longo for useful discussion. 
This work has been partially supported by MURST Cofinanziamento 2000.
This research has made use of \emph{NASA's Astrophysics Data System
Abstract Service}.
\end{acknowledgements}

\end{document}